\documentclass[pre, 10pt, showkeys]{revtex4-2}
\usepackage{etex} 

\usepackage{amsmath, amssymb, commath, braket, bm, empheq, array}

\usepackage{graphicx}
\DeclareGraphicsExtensions{.eps,.jpg,.png,.pdf}
\usepackage{hyperref}
\hypersetup{
	colorlinks=true,
	linkcolor=red,
	filecolor=blue,
	urlcolor=magenta,
}

\usepackage{dblfloatfix}

\usepackage{booktabs, multirow}
\usepackage[inline]{enumitem}
\newcommand\x{1.0}
\newcommand{\adj}{\ensuremath{\mathbb{A}}}	%	Adjacency Matrix
\providecommand{\ceil}[1]{\left \lceil #1 \right \rceil }
\def\cp{\ensuremath\mathcal{P}}	%	Measure of Parity
\def\cc{\ensuremath\mathcal{C}}	%	Measure of Chirality
\def\cf{\ensuremath\mathcal{F}}	%	Fraction of mixed states
\newcommand{\ch}{\ensuremath{\mathcal{H}}}	%	Hamiltonian
\newcommand{\ferf}[1]{\ensuremath{\text{Erf}\del{#1}}}
\newcommand{\fgamma}[1]{\ensuremath{\Gamma\del{#1}}}
\newcommand{\fnrm}[1]{\ensuremath{||#1||_\text{F}}}
\def\graph{\ensuremath \mathcal{G}}					%	Graph
\newcommand{\hgoe}{\ensuremath{H_{\text{GOE}}}}	%	GOE matrix
\def\J{\ensuremath \mathbb{J}}			%	Exchange matrix
\def\jk{\ensuremath{ \langle\mathbb{J}\del{\gamma}\rangle _k } }
\def\mj{\ensuremath \langle\mathbb{J}\rangle}			%	<Exchange matrix>
\newcommand{\mean}[1]{\ensuremath{\overline{#1}}}
\newcommand{\pt}{\ensuremath{\mathbb{P}}}	%	P Matrix
\newcommand{\prob}[1]{\ensuremath{\text{P}\del{#1}}}

\newcommand{\qt}{\ensuremath{\mathbb{Q}}}	%	Q Matrix
\newcommand{\shn}{\ensuremath{\textrm{S}}}
\newcommand{\Tr}[1]{\ensuremath{\text{Tr}\del{#1}}}
\begin{document}
\title{Transport in Deformed Centrosymmetric Networks}
\author{Adway Kumar Das}\email{akd19rs062@iiserkol.ac.in}
\author{Anandamohan Ghosh}\email{anandamohan@iiserkol.ac.in}
\affiliation{
	Indian Institute of Science Education and Research Kolkata, Mohanpur, 741246 India
}
\date{\today}
%--------------------------------------------------------
%	ABSTRACT
\begin{abstract}
	Centrosymmetry often mediates Perfect State Transfer (PST) in various complex systems ranging from quantum wires to photosynthetic networks. We introduce the Deformed Centrosymmetric Ensemble (DCE) of random matrices, $H\del{\lambda} \equiv H_+ + \lambda H_-$, where $H_+$ is centrosymmetric while $H_-$ is skew-centrosymmetric. The relative strength of the $H_\pm$ prompts the system size scaling of the control parameter as $\lambda = N^{-\frac{\gamma}{2}}$. We propose two quantities, $\cp$ and $\cc$, quantifying centro- and skewcentro-symmetry, respectively, exhibiting second order phase transitions at $\gamma_\text{P}\equiv 1$ and $\gamma_\text{C}\equiv -1$. In addition, DCE posses an ergodic transition at $\gamma_\text{E} \equiv 0$. Thus equipped with a precise control of the extent of centrosymmetry in DCE, we study the manifestation of $\gamma$ on the transport properties of complex networks. We propose that such random networks can be constructed using the eigenvectors of $H\del{\lambda}$ and establish that the maximum transfer fidelity, $F_T$, is equivalent to the degree of centrosymmetry, $\cp$.
\end{abstract}
%--------------------------------------------------------
%{Matrix Theory}
%{Quantum communication}
%{Systems obeying scaling laws}
%--------------------------------------------------------
\maketitle
%=========================================================
%	SECTIONS
%=========================================================
\section{Introduction}
%=========================================================
The ideal choice for communication in quantum devices is a quantum wire, described by a network of $N$ qubits. These systems are experimentally realizable in optical lattices \cite{Mandel1}, quantum dots \cite{Loss1}, NMR \cite{Gershenfeld1}, trapped ions \cite{Brown1}, superconducting qubits \cite{Karamlou1}. Using such quantum wires, one can achieve Perfect State Transfer (PST), i.e.~the transmission of a state between two locations with unit fidelity. This can be implemented using a series of SWAP gates \cite{Kane1} or multiple-spin encoding \cite{Haselgrove1}, however, at the cost of introducing decoherence \cite{Benjamin1}. Contrarily, many pre-engineered quantum wires allow PST without external dynamical control, e.g.~certain spin-chains \cite{Christandl1, Karbach1, Shi1, Venuti1}, arrays of quantum dots \cite{Nikolopoulos1} with wide range of potential applications from mathematical finance \cite{Albanese1} to biology \cite{Lambert1}.

Surprisingly, all the above systems exhibiting PST posses centrosymmetry (also known as exchange/ mirror-symmetry). The degree of centrosymmetry is highly correlated to the transfer fidelity in the photosynthetic structures \cite{Scholak1, Zech1} and disordered networks modeled by embedded ensembles \cite{Ortega1}. Centrosymmetry is interesting on its own right and holds a special place in diverse disciplines ranging from information theory \cite{Gersho1, Magee1} to engineering problems \cite{Datta1, Gaudreau1}. Schematic of a typical centrosymmetric matrix is shown in Fig.~\ref{fig_mat_str}(a). Imposing diagonal symmetry results in additional correlations and produces the well known Symmetric Centrosymmetric (SC) matrices \cite{Collar1, Cantoni1}, as depicted in Fig.~\ref{fig_mat_str}(b). In this work we construct an ensemble of random SC matrices where the centrosymmetry can be broken by a skew-centrosymmetric perturbation. Fig.~\ref{fig_mat_str}(c) shows the schematic of a typical Symmetric Skew-Centrosymmetric (SSC) matrix \cite{Abu-Jeib1}, having the chiral symmetry, which is an important ingredient in topological superconductors \cite{Beenakker1}, spin-liquids \cite{Schneider1}, QCD \cite{Verbaarschot1} and experimentally realized in microwave platforms \cite{Rehemanjiang1}. Here we achieve a precise control of the extent of centrosymmetry in a Hamiltonian and study its dynamical manifestations. We demonstrate that the transport in a random network is enhanced as the extent of centrosymmetry is increased.

The organization of the current paper is as follows: in Sec.~\ref{sec_model}, we construct the Deformed Centrosymmetric Ensemble (DCE) parametrized by $\lambda$ using the maximum entropy principle and obtain the density of matrices. The competition between centrosymmetry preserving and breaking parts of the DCE matrices prompts the system size scaling of the control parameter as $\lambda = N^{-\frac{\gamma}{2}}$. In Sec.~\ref{sec_phase}, we compute the expectation value of the exchange operator and propose measures of centrosymmetry ($\cp$) and skew-centrosymmetry ($\cc$). We classify the energy states of DCE according to their parity, i.e.~the change of sign under the action of exchange operator. We observe second order phase transitions in the above measures at three critical points, viz. $\gamma = \pm 1, 0$, separating four distinct phases in DCE. In Sec.~\ref{sec_pst}, we construct various adjacency matrices allowing PST and analytically compute the time evolution of the fidelity of state transfer in such systems. In Sec.~\ref{sec_fidelity}, we discuss the construction of quantum wires allowing tunable transport and show that the maximum fidelity obtainable is related to the measure of the centrosymmetry. Finally our concluding remarks are given in Sec.~\ref{sec_conclusion}.
%=========================================================
\renewcommand{\x}{0.5}
\begin{figure}[t]
	\vspace{-10pt}
	\includegraphics[width=\x\textwidth, trim=0 355 0 335, clip=false]{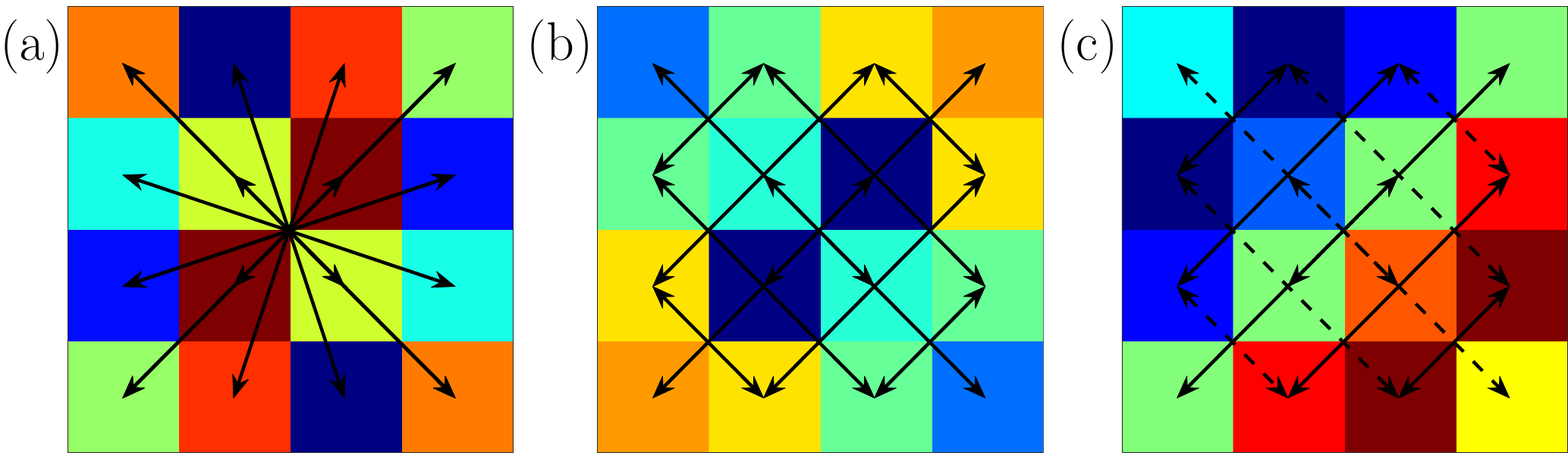}
	\caption{\textbf{Schematic representations} of (a)~{Centrosymmetric,} (b)~{Symmetric Centrosymmetric (SC)} (c)~{Symmetric Skew-centrosymmetric (SSC)} matrices. Solid lines show correlation, while the dashed lines show anti-correlation among the matrix elements.
	}
	\label{fig_mat_str}
\end{figure}
%=========================================================
%=========================================================
\section{Deformed Centrosymmetric Ensemble}\label{sec_model}
%=========================================================
Let, $\J\equiv \cbr{\delta_{i, N+1-j}}$ be the exchange matrix. Then any real $N\times N$ matrix, $H$, can be decomposed into a centrosymmetric ($\ch_+$; $\sbr{\ch_+, \J} = 0$) and a skew-centrosymmetric ($\ch_-$; $\cbr{\ch_-, \J} = 0$) matrix as
\begin{align}
	\label{eq_H_decomp}
	H &= \ch_+ + \ch_-,\quad \ch_\pm = \frac{H\pm \J H\J}{2}.
\end{align}
Some of the general properties of $\ch_\pm$ are discussed in Appendices~\ref{sec_rsc} and \ref{sec_rss}. Since $\J $ is invertible, the decomposition in Eq.~\eqref{eq_H_decomp} is always possible. The squared Frobenius norm of the components of $H$ can be expressed as $\fnrm{\ch_\pm}^2 = \frac{1}{2}\del{\Tr{H^2} \pm \Tr{\del{\J H}^2}}$, which we constrain to understand the statistical properties of $H$. We find the density, $\prob{H}$, satisfying the following properties: $\prob{H}$ must be normalized; the squared norm, $\Tr{H^2}$, is finite; the extent of centrosymmetry, $\Tr{\del{\J H}^2}$, is equal on average for all the members of the ensemble. Defining the expectation value as $\mean{\mathcal{O}} = \int dH\prob{H}\mathcal{O}$, the above conditions can be given as
\begin{align}
	\label{eq_H_cen_constr}
	\mean{\mathbb{I}} = 1,\quad \mean{\Tr{H^2}} = \mu,\quad \mean{\Tr{\del{\J H}^2}} = \nu.
\end{align}
The relative strength of $\mu$ and $\nu$ dictates the competition between $\ch_\pm$ and can be tuned to control the extent of centrosymmetry in $H$. The skew-centrosymmetric perturbation from $\ch_-$ deforms the correlation present among the matrix elements of $\ch_+$ and $H$ is defined to belong to the Deformed Centrosymmetric Ensemble (DCE). Our matrix model is closely related to the Deformed Gaussian Orthogonal Ensemble (DGOE) \cite{Hussein2} and can be obtained by a similarity transformation by the eigenbasis of the exchange matrix (see Appendix~\ref{sec_dgoe}).

We now use the Maximum Entropy Principle (MEP) to obtain $\prob{H}$ in DCE analogous to the calculations for DGOE \cite{Hussein2} and Deformed Poisson Ensemble \cite{Das3}. We maximize the Shannon entropy $\shn = -\int dH \prob{H}\ln\prob{H}$ using variational principle subjected to the constraints in Eq.~\eqref{eq_H_cen_constr}, implying $\delta\sbr{\shn - \zeta\mean{\mathbb{I}} - \alpha\mean{\Tr{H^2}} - \beta\mean{\Tr{\del{\J H}^2}}} = 0$. Consequently the density of matrices in DCE can be expressed in terms of the Lagrange multipliers $\alpha, \beta, \zeta$ as
\begin{align}
	\label{eq_P_H}
	\prob{H} = \dfrac{1}{Z}\exp\del{-\alpha\Tr{H^2} - \beta\Tr{\del{\J H}^2}}
\end{align}
where $Z = e^{1+\zeta}$ is the normalization constant. Solving the integrals in Eq.~\eqref{eq_H_cen_constr} we get
\begin{align}
	\label{eq_integrals}
	\begin{split}
		Z &= \dfrac{\pi^{ \frac{N(N+1)}{4} }}{2^{\frac{N(N-1)}{4}}\del{\alpha^2 - \beta^2}^{\frac{N^2}{8}}\del{\alpha + \beta}^{\frac{N}{4}}}\\
		\mu &= \frac{N}{4(\alpha + \beta)}\del{1 + \frac{N\alpha}{\alpha - \beta}},\: \nu = \frac{N}{4(\alpha + \beta)}\del{1 - \frac{N\beta}{\alpha - \beta}}.
	\end{split}
\end{align}
Then the Lagrange multipliers, $\alpha$ and $\beta$, can be expressed in terms of the constraints ($\mu, \nu$) for $N\gg 1$
\begin{align}
	\label{eq_a_b}
	\alpha\approx \dfrac{N^2\mu}{4(\mu^2 - \nu^2)},\quad \beta\approx \dfrac{-N^2\nu}{4(\mu^2 - \nu^2)}
\end{align}
%=========================================================
\begin{table}[t]
	%	\centering
	\begin{tabular}{|c|c|c|c|c|c|}
		\hline
		$H$	& $\lambda$	& $\beta$ 			& $\gamma$	& $\prob{\mj}$ \\ \hline
		SSC	& $\infty$	& -$\dfrac{1}{2}$	& $-\infty$ & $\delta\del{0}$\\ \hline
		GOE	& 1			& 0 				& 0 		& $\mathcal{N}\del{0, \dfrac{2}{N}}$\\ \hline
		SC	& 0			& $\dfrac{1}{2}$ 	& $\infty$ 	& $\dfrac{\delta\del{-1} + \delta\del{1}}{2}$\\ \hline
	\end{tabular}
	\caption{Limiting cases of DCE taking $\alpha = \frac{1}{2}$. Different parametrizations of DCE are related as: $\beta = \frac{1-\lambda^2}{1+\lambda^2}$ and $\lambda = N^{-\frac{\gamma}{2}}$. $\prob{\J}$ is the density of the expectation value of the exchange operator, $\J$.}
	\label{tab_H_limits}
\end{table}
%=========================================================
giving a complete statistical description of $H$. The relative strength of $\ch_\pm$ can be expressed as a function of the Lagrange multipliers, $\alpha, \beta$:
\begin{align*}
	\frac{\fnrm{\ch_+}^2}{\fnrm{\ch_-}^2} = \frac{(N+2)(\alpha - \beta)}{N(\alpha + \beta)} \approx \frac{\alpha - \beta}{\alpha + \beta}\equiv \lambda^2
\end{align*}
giving an equivalent matrix model
\begin{align}
	\label{eq_H_model_J}
	H(\lambda) &= H_+ + \lambda H_-,\quad H_\pm = \dfrac{\hgoe \pm \J \hgoe \J }{2}.
\end{align}
Here $H_\pm$ are centrosymmetric and skew-centrosymmetric, respectively, and $\hgoe\in$ GOE is a symmetric random matrix with Gaussian distributed elements \cite{Mehta1}. Since $\alpha$ can be eliminated by appropriately setting the energy scale, we choose $\alpha = \frac{1}{2}$. Then the density of matrices from Eq.~\eqref{eq_P_H} can be expressed as,
\begin{align}
	\label{eq_P_H_modified}
	\begin{split}
		\prob{H(\lambda)} &= \frac{1}{Z}\exp\del{-\frac{1}{2}\Tr{H^2} - \frac{1-\lambda^2}{ 2\del{1+\lambda^2} }\Tr{\del{\J H}^2}}\\
		Z &= \frac{ \del{\pi\del{1 + \lambda^2}}^{\frac{N(N+1)}{4}} }{ 2^{\frac{N(N-1)}{4}}\lambda^{\frac{N^2}{4}} }
	\end{split}
\end{align}
Equipped with the analytical expression of $\prob{H(\lambda)}$, it is now possible to calculate the Shannon entropy, $\shn(\lambda)$, characterizing the extent of correlations in a random matrix in DCE (explicit calculations shown in Appendix~\ref{sec_moment}). We tune $\lambda$ such that $H(\lambda)$ exhibits different symmetric/chiral phases, which will be characterized in the following section.
%=========================================================
%=========================================================
\section{Phase Transitions in DCE}\label{sec_phase}
%=========================================================
All the elements of $H_\pm$ follow normal distribution, hence their typical fluctuations, $\Delta H_\pm(i, j)\sim \mathcal{O}(1)$. The coordination numbers of the diagonal elements for both the matrices are $\mathcal{O}\del{N}$. For $\lambda = 1$, the equal contribution from $H_+$ and $\lambda H_-$ completely breaks the centrosymmetry of $H(\lambda)$ in Eq.~\eqref{eq_H_model_J} and produces a GOE matrix. $H(\lambda = 1)$ belonging to GOE is reminiscent of the emergence of GOE in the Rosenzweig-Porter ensemble \cite{Kravtsov1, Das1}. Moreover, due to the random sign altering nature of the elements of DCE matrices, there exists another coordination number $\sim\mathcal{O}\del{\sqrt{N}}$. Thus the critical behaviors can be expected for $\sqrt{N}\Delta H_+ \sim \lambda N\Delta H_-\Rightarrow \lambda \sim \frac{1}{\sqrt{N}}$ and $N\Delta H_+ \sim \lambda \sqrt{N}\Delta H_-\Rightarrow \lambda \sim \sqrt{N}$. Thus we consider the following system size scaling
\begin{align}
	\label{eq_lambda}
	\lambda = N^{-\frac{\gamma}{2}},\qquad \gamma\in\mathbb{R}.
\end{align}
Such scaling suggests possible criticalities at $\gamma = \pm 1, 0$, separating four distinct phases based on their symmetry/ chiral properties, which can be identified by the following measures.
%=========================================================
\renewcommand{\x}{0.5}
\begin{figure}[t]
	\vspace{-10pt}
	\includegraphics[width=\x\textwidth, trim=0 295 0 265, clip=false]{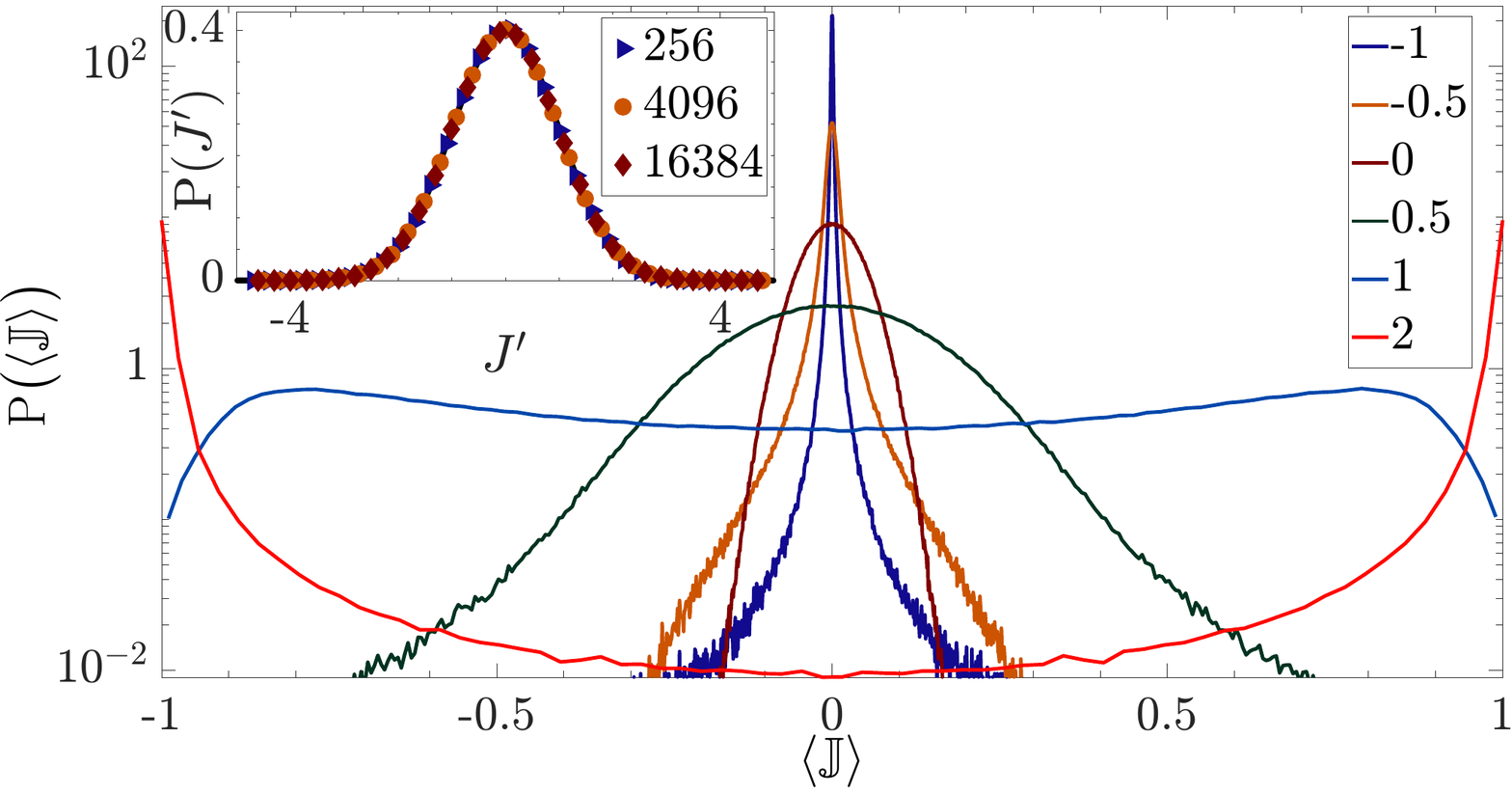}
	\caption[]{Density of the expectation value of the exchange operator, $\J$, for various $\gamma$ and $N = 1024$. % ($\lambda = N^{-\frac{\gamma}{2}}$). 
		Inset shows density of $J' = \sqrt{\frac{N}{2}}\mj$ for $\gamma = 0$ %(i.e. $\lambda = 1$) 
		and various system sizes, $N$, where the bold line denotes normal distribution, $\mathcal{N}\del{0, 1}$.
	}
	\label{fig_2}
\end{figure}
%=========================================================
\paragraph*{\bf Measure of centrosymmetry:} In our matrix model (Eq.~\eqref{eq_H_model_J}), $\lambda$ controls the extent of centrosymmetry, reflected in $\mj\equiv \bra{\Psi}\J \ket{\Psi}$, where $\ket{\Psi}$ is any wavefunction. In SC matrices, $\mj$ can be 1 or -1 as corresponding energy states are symmetric ($\J\ket{\Psi} = \ket{\Psi}$) or antisymmetric ($\J\ket{\Psi} = -\ket{\Psi}$). Contrarily for SSC matrices, if $\del{E, \ket{\Psi}}$ is an eigenpair, then so is $\del{-E, \J \ket{\Psi}}$, hence $\ket{\Psi}$ is orthogonal to $\J \ket{\Psi}$ and $\mj = 0$. In the case of equal contributions from SC and SSC matrices, i.e.~for $H\del{\lambda = 1}\in$ GOE, the eigenvector components can be regarded as {i.i.d.}~random variables \cite{Berry5}, implying $\mj$ follows $\mathcal{N}\del{0, \frac{2}{N}}$, i.e.~Gaussian distribution with variance $\frac{2}{N}$. In Fig.~\ref{fig_2}, we show the density, $\prob{\mj}$, for various $\gamma$ values for matrix size $N = 1024$. For $\gamma\ll 0$, $\prob{\mj}$ is sharply peaked at 0, while for $\gamma\gg 0$, the distribution is bimodal with peaks at $\pm 1$ and $\prob{\mj}_{\gamma = 0}\sim \mathcal{N}\del{0, \frac{2}{N}}$.
%=======================================================
Since $\prob{\mj}$ is symmetric about zero, the ensemble average of $\mj$ is 0 $\forall\:\gamma$. Thus to measure the presence of centrosymmetry, we look at the quantity,
\begin{align}
	\label{eq_def_P}
	\cp = \abs{\mj}^2 = \abs{\bra{\Psi} \J \ket{\Psi}}^2
\end{align}
Clearly $\cp = 1$ and 0 for centrosymmetric and skew-centrosymmetric matrices, respectively, while we get the Gamma distribution $\fgamma{\frac{1}{2}, \frac{4}{N}}$ for the GOE limit. In Fig.~\ref{fig_3}(c), we show the ensemble average, $\mean{\cp}$, as a function of $\gamma$ for different system sizes, $N$. We observe that for $\gamma\gg 1$ ($\lambda\to 0$) $H$ becomes a SC matrix giving $\mean{\cp}\to 1$ independent of the system size, $N$. Similarly $H$ is a SSC matrix and $\mean{\cp}\to 0$ for $\gamma\ll -1$. Exactly at $\gamma = 0$, $\mean{\cp} = \frac{2}{N}$, which is the mean of the distribution $\fgamma{\frac{1}{2}, \frac{4}{N}}$ and we find that $\mean{\cp}\propto N^{\gamma-1}$ for $0 \leq \gamma\leq 1$. We numerically observe that $\mean{\cp}_{\gamma = 1} \approx 0.3$, independent of the system size. Consequently the crossover curves in Fig.~\ref{fig_3}(c) exhibit a non-analyticity around $\gamma = 1$ and we are able to collapse the data for different system sizes, as shown in the inset of Fig.~\ref{fig_3}(c). This confirms that $\cp$ undergoes a second order phase transition at $\gamma_\text{P} \equiv 1$ and DCE matrices are centrosymmetric for $\gamma>1$ in the thermodynamic limit.
%==========================================
\paragraph*{\bf Identification of mixed states:} So far we have observed that $\overline{\cp}<1$ for $\gamma<1$, hence the corresponding eigenfunctions are neither strictly odd nor even. Such distinction is completely lost in the case of GOE. At $\gamma = 0$, DCE matrices belong to GOE, such that $\cf_{\gamma = 0} = \ferf{\frac{z}{\sqrt{2}}}$ is the fraction of states with $\mj$ in the interval $\Delta\J\equiv \del{ -\sqrt{\frac{2}{N}}z, \sqrt{\frac{2}{N}}z }$, i.e.~$z$-$\sigma$ confidence interval of the Gaussian distribution $\mathcal{N}\del{0, \frac{2}{N}}$. Hence for a general Hamiltonian, we identify any eigenvector with $\mj\in\Delta\J$ to be a mixed wavefunction, which has no parity at all. In Fig.~\ref{fig_3}(b), we show the ensemble average of the fraction of mixed states, $\mean{\cf}$ vs.~$\gamma$ for different $N$ taking $z=1$ and get $\mean{\cf}_{\gamma = 0} \approx 0.68 \approx \ferf{\frac{1}{\sqrt{2}}}$ in agreement with GOE behavior. In general, $\mean{\cf}$  shows a second order transition at $\gamma=0$. Consequently for $\gamma\leq \gamma_\text{E} \equiv 0$, all the energy states of DCE are ergodic as in GOE, else the ergodicity is lost as the extent of centrosymmetry increases.
%%=========================================================
\paragraph*{\bf Chirality as a measure of skew-centrosymmetry:} In Eq.~\ref{eq_H_model_J}, $\lambda\gg 1$ implies that the contribution from $H_+$ can be ignored and effectively $\cbr{H, \J} = 0$. Since $\del{E, \ket{\Psi}}$ and $\del{-E, \J\ket{\Psi}}$ both are eigenpairs of SSC matrices \cite{Hill1}, we propose the following measure of chirality,
\begin{align}
	\label{eq_def_C}
	\cc = \abs{\bra{\Psi_i} \J \ket{\Psi_{i'}}}^2,\quad i' \equiv N+1-i
\end{align}
where $\Ket{\Psi_i}$ is the $i^{th}$ eigenstate (sorted in the ascending order of eigenvalues) of an $N\times N$ matrix. Clearly $\cc = 1$ for SSC matrices and $\cc = 0$ for SC matrices. In the case of GOE, $\prob{\cc_\text{GOE}}\sim\fgamma{\dfrac{1}{2}, \dfrac{2}{N}}$ with mean $\dfrac{1}{N}$, hence $\cc_\text{GOE}\to 0$ for $N\to\infty$, indicating the absence of chiral symmetry. For $\gamma\ll -1$,  $H$ becomes a SSC matrix giving $\mean{\cc}\to 1$ independent of $N$.
Again for $\gamma\gg 1$, $H_+$ dominates and skew-centrosymmetry is completely lost such that $\mean{\cc}\to 0$.  We also observe that $\mean{\cc}\propto N^{-\gamma-1}$ for $-1\leq \gamma\leq 0$. Scaling analysis suggests a second order transition at $\gamma_\text{C} \equiv -1$ where we obtain $\mean{\cc}\approx 0.25$ independent of the system size.

The measure of centro- and chiral symmetry, ($\cp$ and $\cc$) along with the fraction of mixed states, $\cf$, indicate the existence of three second order phase transitions in DCE at $\gamma = \pm 1, 0$, respectively, separating four distinct phases (Fig.~\ref{fig_3}). The role of centro- and chiral symmetries in transport are studied in the subsequent sections.
%=========================================================
\renewcommand{\x}{1}
\begin{figure*}[t!]
	\vspace{-10pt}
	\includegraphics[width=\x\textwidth, trim=0 330 0 315, clip=false]{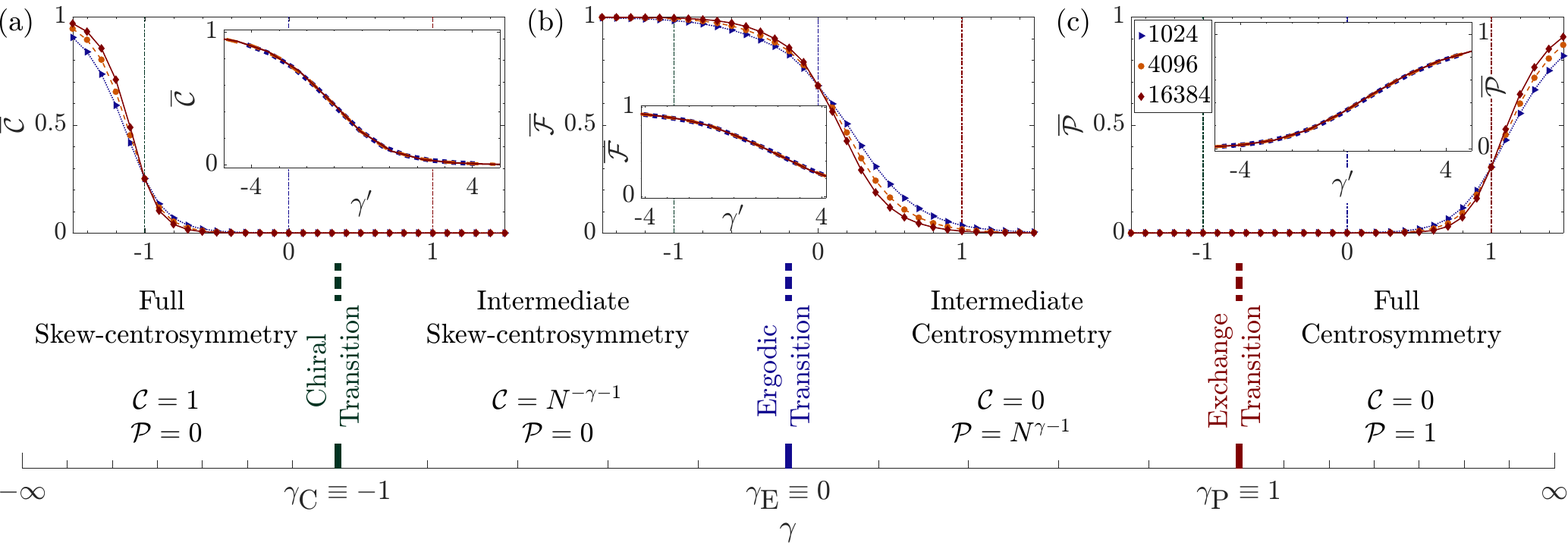}
	\caption[]{Ensemble average of various quantities for different system sizes, $N$, as a function of $\gamma$: (a) $\cc$, the measure of chiral symmetry, (b) $\cf$, the fraction of $\ket{\Psi}_\text{mixed}$ taking $z = 1$ in $\Delta\J$ and (c) $\cp$, the measure of centrosymmetry. Insets show collapsed data using second order transition ansatz \cite{Das2}, where $\gamma' = (\gamma-\bar{\gamma})\del{\ln N}^\frac{1}{\nu}$, $\bar{\gamma}$ = critical point and $\nu$ = critical exponent. For the three quantities investigated, the critical points are $-0.99641, -0.00902$ and $1.00776$ along with the critical exponents $1.09164, 1.08806$ and 1.03264, respectively. Thus three critical points separate four distinct phases as shown above.
	}
	\label{fig_3}
\end{figure*}
%=========================================================
%=========================================================
\section{Perfect Transport}\label{sec_pst}
%=========================================================
Let, $\graph$ be a simple undirected graph with vertices labelled as $1,2,\dots, N$. Corresponding adjacency matrix, $\adj$, with an eigendecomposition  $\cbr{E_k, \ket{\Phi_k}}$ can be expressed as
\begin{align}
	\label{eq_Adj}
	\begin{split}
		\adj &= \sum_{k = 1}^{N}E_k\ket{\Phi_k}\bra{\Phi_k}\\
		&\equiv \sum_{j = 1}^{N/2} E_{2j-1}\ket{\Phi_{2j-1}}\bra{\Phi_{2j-1}} + E_{2j}\ket{\Phi_{2j}}\bra{\Phi_{2j}}.
	\end{split}
\end{align}
Now we demand that $\adj$ has centrosymmetry as in $H_+$, such that, $\ket{\Phi_{2j-1}} = \frac{1}{\sqrt{2}}\begin{pmatrix}
	\ket{y_j}\\
	-\J \ket{y_j}
\end{pmatrix}$ and $\ket{\Phi_{2j}} = \frac{1}{\sqrt{2}}\begin{pmatrix}
	\ket{u_j}\\
	\J \ket{u_j}
\end{pmatrix}$, where $\cbr{\ket{y_j}}$ and $\cbr{\ket{u_j}}$ are orthonormal bases. Given such a centrosymmetric adjacency matrix, we want to transfer a state initially localized at vertex 1 to the vertex $N$ of the graph $\graph$. The fidelity of such an excitation transfer has the time evolution,
\begin{align}
	\label{eq_fidelity}
	F(t) = \abs{ \bra{N}e^{-i\adj t}\ket{1} }^2
\end{align}
where $\ket{j} \equiv \hat{e}_j$. A Perfect State Transfer (PST) implies that $F_T \equiv F(t = T) = 1$. Note that PST is not realizable in a generic random SC matrix, as an initially localized state diffuses, while the network equilibrates to a state with the initial excitation uniformly spread over all the vertices. Nevertheless, the periodicity of $F_T$ along with the orthonormality of the eigenbasis, $\cbr{\Phi_k}$, lead to the following {\it necessary and sufficient} conditions for PST in any centrosymmetric $\adj$,
\begin{align}
	\label{eq_pst_constraint}
	\begin{split}
		\sum_{k = 1}^{N} (-1)^kr_k e^{-i E_k T} &= 1\\
		\sum_{k = 1}^{N}(-1)^kr_k &= 0, \; 2T \textrm{ Periodic}\\
		\sum_{k = 1}^{N}r_k &= 1, \; \textrm{Orthonormality}
	\end{split}
\end{align}
where $r_k \equiv \Phi_k(1)^2\in[0, 1]$. Here we ignore any global phase factor in the time evolved state at $t=T$, which can be nullified by an overall shift of the energy axis. The criteria in Eq.~\eqref{eq_pst_constraint} are equivalently known as the {\it spectrum parity matching conditions} \cite{Shi1}.

There exist various solutions of Eq.~\eqref{eq_pst_constraint}, where both the energy levels and the eigenstates of $\adj$ i.e.~all the $E_k$'s and $r_k$'s are constrained \cite{Albanese1, Christandl4}. However, Eq.~\eqref{eq_pst_constraint} implies that out of $2N$ variables, we have the freedom to constrain any $N$ variables. We now illustrate two cases where either $E_k$'s or $r_k$'s are constrained.
%=======================================================================
\paragraph*{\bf Example 1:} One of the simplest solutions of Eq.~\eqref{eq_pst_constraint} is to consider any random energy sequence $\cbr{E_k}$, while both $\cbr{\ket{y_j}}$ and $\cbr{\ket{u_j}}$ are unit bases, leading to the eigenvectors,
\begin{align}
	\label{eq_A_ladder_vec}
	\ket{\Phi_k} &= \frac{1}{\sqrt{2}}\del{\dots,0, \underset{\ceil{\frac{k}{2}}}{\underbrace{(-1)^k}}, 0\dots,0, \underset{N+1-\ceil{\frac{k}{2}}}{\underbrace{1}}, 0,\dots}.
\end{align}
Then the adjacency matrix can be expressed as a superposition of diagonal and anti-diagonal elements,
\begin{align}
	\label{eq_A_ladder}
	\begin{split}
		\adj &= \frac{1}{2}\begin{pmatrix}
			A& C\J \\
			\J C& \J A\J 
		\end{pmatrix},\quad \begin{matrix}
			A_{ij} = \delta_{ij}(E_{2j} + E_{2j-1})\\
			C_{ij} = \delta_{ij}(E_{2j} - E_{2j-1})
		\end{matrix}.
	\end{split}
\end{align}
For spin-1/2 particles with XY interaction, $\adj$ represents an abstract spin-ladder \cite{Song1} of length $\frac{N}{2}$: there is no interaction among the spins of individual chain while the tunneling amplitude between $i^{th}$ sites of chain 1 and 2 is $\frac{A_{ii}}{2}$ and both the sites are under an external magnetic field of strength $\frac{C_{ii}}{2}$. In this setup, the dynamics is trivial as the first vertex is coupled only to the $N^{th}$ vertex and the fidelity of transport can be exactly calculated,
\begin{align}
	\label{eq_fidelity_ladder}
	\begin{split}
		\bra{1}e^{-i\adj t}\ket{N} &= \sum_{j = 1}^{N}e^{-iE_jt}\Phi_j(1)\Phi_j(N)\\
		&= i\exp\del{-i\frac{E_2+E_1}{2}t}\sin\del{\frac{E_2-E_1}{2}t}\\
		\Rightarrow F(t) &= \abs{\bra{1}e^{-i\adj t}\ket{N}}^2 = \sin^2\del{\frac{E_2-E_1}{2}t}
	\end{split}
\end{align}
with PST at $T = \dfrac{\pi}{E_2 - E_1}$. $F(t)$ can be numerically evaluated using Eq.~\eqref{eq_fidelity} by constructing $\adj$ as in Eq.~\eqref{eq_A_ladder} from uniformly distributed random numbers as the energy levels. The resultant time evolution is shown in Fig.~\ref{fig_4}(a) for different $N$, in agreement with the expression in Eq.~\eqref{eq_fidelity_ladder}. 
%=========================================================
\renewcommand{\x}{1}
\begin{figure}[t]
	\vspace{-10pt}
	\includegraphics[width=\x\textwidth, trim=0 378 0 365, clip=false]{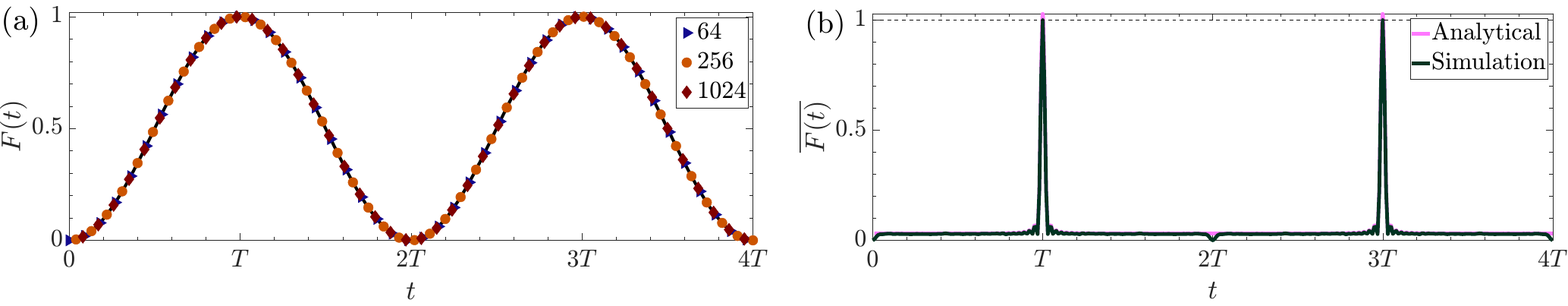}
	\caption[]{{\bf Time evolution of the transfer fidelity} for the adjacency matrix in (a) Eq.~\eqref{eq_A_ladder}. The solid line represents Eq.~\eqref{eq_fidelity_ladder}, while the markers denote numerical simulations for different number of vertices, $N$, and arbitrary energy spectra, where $T = \frac{\pi}{E_2 - E_1}$, $E_{1,2}$ being the first two energy levels.
		%=============================
		(b) Fidelity of state transfer averaged over $10^3$ realizations of the $64\times 64$ adjacency matrices in Eq.~\eqref{eq_Adj} with the energy levels $\vec{E} = \cbr{1, 2,\dots,64}$, while $\ket{y_j}$ and $\ket{u_j}$'s are random vectors. The bold curve indicates the approximate analytical form (Eq.~\eqref{eq_fidelity_N}).
	}
	\label{fig_4}
\end{figure}
%==========================================================
\paragraph*{\bf Example 2:} Now we consider $\ket{y_j}$'s and $\ket{u_j}$'s in Eq.~\eqref{eq_Adj} to be random vectors similar to those of GOE. Thus $r_k > 0$ in general and the energy levels must be constrained to satisfy Eq.~\eqref{eq_pst_constraint}. Since we ignore any global phase acquired after time $T$, the only solution of Eq.~\eqref{eq_pst_constraint} is
\begin{align}
	\label{eq_pst_energy}
	E_k = \frac{\pi}{T}\del{k(2q_k + 1) - 2p_k},\quad k = 1,2,\dots,N
\end{align}
where $q_k, p_k$ are arbitrary integers. Eq.~\eqref{eq_pst_energy} is a generalization of the linear spectrum of Krawtchouk chain \cite{Albanese1}. The emergence of PST  becomes obvious as the condition Eq.~\eqref{eq_pst_energy} ensures that the time propagator, $e^{-i\adj t}$, becomes the exchange matrix, $\J$, at $t = T$. 
The choice of arbitrary orthonormal bases $\cbr{\ket{y_j}}$ and $\cbr{\ket{u_j}}$ in the spectral decomposition of Eq.~\eqref{eq_Adj} is complimentary to the schemes proposed in \cite{Yung1, Kostak1} and allows the construction of infinitely many adjacency matrices allowing PST. Using Eq.~\eqref{eq_fidelity}, we numerically compute the fidelity averaged over an ensemble of random $\adj$ exhibiting PST as shown in Fig.~\ref{fig_4}(b).

An exact calculation is difficult but an approximate expression can be obtained  by considering a linear spectrum, $\cbr{E_k = k}$. Then the fidelity in Eq.~\eqref{eq_fidelity} can be expressed as,
\begin{align}
	\label{eq_fidelity_t_linear}
	F(t) = \sum_{k = 1}^{N}r_k^2 + 2\sum_{k = 1}^{N-1}(-1)^k\cos\del{kt}\sum_{j = 1}^{N-k}r_jr_{j+k}.
\end{align}
We can assume that $x \equiv Nr_k$ follows the Porter-Thomas distribution, $\prob{x} = \del{\sqrt{2\pi x}e^{\frac{x}{2}}}^{-1}$ since $\ket{y_j}$'s and $\ket{u_j}$'s in Eq.~\eqref{eq_Adj} are random vectors \cite{Luca1}. Then the ensemble averaged fidelity of $\adj$ becomes
\begin{align}
	\label{eq_fidelity_N}
	\overline{F(t)} \sim \frac{2}{N} + \frac{1}{N^2}\sin^2\frac{Nt}{2}\sec^2\frac{t}{2}.
\end{align}
Note that $F(t\to 0)\to \frac{2}{N}$ and $F(t\to\pi)\to 1+\frac{2}{N}$, which in the thermodynamic limit converge to 0 and 1, satisfying the constraints of periodic dynamics and orthonormality given in Eq.~\eqref{eq_pst_constraint}. In Fig.~\ref{fig_4}(b), we show the expression in Eq.~\eqref{eq_fidelity_N} for $N = 64$ along with the numerical simulation.
%=========================================================
%=========================================================
\section{Transport using DCE}\label{sec_fidelity}
%=========================================================
Now we want to control the transfer fidelity in a graph $\graph$ by tuning the extent of centrosymmetry using $\lambda = N^{-\frac{\gamma}{2}}$. Here also we construct an adjacency matrix with $E_k$'s satisfying Eq.~\eqref{eq_pst_energy} and $\cbr{\ket{\Phi_k^{\del{\gamma}}}}$, the eigenbasis of $H(\lambda)\in$ DCE:
\begin{align}
	\label{eq_tunable_A}
	\begin{split}
		\adj\del{\gamma} &= \sum_{k = 1}^{N} E_k \ket{\Phi_k^{\del{\gamma}}} \bra{\Phi_k^{\del{\gamma}}}\\
%		&\equiv \sum_{j = 1}^{N/2} E_{2j-1}\ket{\Phi_{2j-1}^{\del{\gamma}}}\bra{\Phi_{2j-1}^{\del{\gamma}}} + E_{2j}\ket{\Phi_{2j}^{\del{\gamma}}}\bra{\Phi_{2j}^{\del{\gamma}}}.
	\end{split}
\end{align}
We define $\jk\equiv \bra{\Phi_k^{\del{\gamma}}}\J \ket{\Phi_k^{\del{\gamma}}}$ and $\ket{ \Phi_k^{\del{\gamma}} }$'s are ordered in such a way that $\langle \J\del{\gamma}\rangle_{2j-1}<0$ and $\langle \J\del{\gamma}\rangle_{2j}>0$, i.e. $(-1)^k\jk = \abs{\jk}$. This construction ensures PST for $\gamma>1$ as $\adj\del{\gamma}$ becomes centrosymmetric for large $N$, while for $\gamma<0$, there is no transport. In Fig.~\ref{fig_5}(a), we show $F(t)$ for different $\gamma$ averaged over $10^4$ disordered realizations of $16\times 16$ $\adj\del{\gamma}$. We observe that the maximum fidelity decreases with $\gamma$ and always occurs at $T$. This is expected as lowering $\gamma$ breaks the centrosymmetry in $\adj\del{\gamma}$ while the energy levels are independent of $\gamma$. Now we want to calculate the ensemble averaged maximum transport fidelity and show that it is equivalent to the extent of centrosymmetry in $\adj\del{\gamma}$.

Let us begin with the decomposition $\ket{\Phi_k^{\del{\gamma}}} = \ket{\Phi_{k+}^{\del{\gamma}}} + \ket{\Phi_{k-}^{\del{\gamma}}},\: \ket{\Phi_{k\pm}^{\del{\gamma}}} = \frac{\ket{\Phi_k^{\del{\gamma}}} \pm \J \ket{\Phi_k^{\del{\gamma}}}}{2}$, where the relative strength of the two components can be expressed as,
\begin{align}
	\label{eq_DCE_V_decomp}
	\dfrac{\fnrm{\Phi_{k+}^{\del{\gamma}}}}{\fnrm{\Phi_{k-}^{\del{\gamma}}}} = \sqrt{\dfrac{1 + \jk}{1 - \jk}}.
\end{align}
Then using the normalization of $\ket{\Phi_k^{\del{\gamma}}}$, we get
\begin{align}
	\ket{\Phi_k^{\del{\gamma}}} &= \sqrt{\frac{1 + \jk}{2}}\ket{\Phi^+} + \sqrt{\frac{1 - \jk}{2}}\ket{\Phi^-}
\end{align}
where $\ket{\Phi^\pm}$ are normalized even and odd wavefunctions. The randomness of DCE allows us to assume that $\Phi^\pm(i)$ follow the Gaussian distribution, $\mathcal{N}\del{0, \frac{1}{N}}\;\forall\; i$. Then it is straightforward to see that
\begin{align}
	\label{eq_state_decomp}
	\begin{split}
		\Phi_k^{\del{\gamma}}(1)\Phi_k^{\del{\gamma}}(N) &= \frac{\Phi^+(1)^2-\Phi^-(1)^2}{2}\\&+\jk\frac{\Phi^+(1)^2 + \Phi^-(1)^2}{2}\\
		\Rightarrow \overline{\Phi_k^{\del{\gamma}}(1)\Phi_k^{\del{\gamma}}(N)} &= \frac{\jk}{N}
	\end{split}
\end{align}
%=========================================================
\renewcommand{\x}{1}
\begin{figure}[t]
	\vspace{-10pt}
	\includegraphics[width=\x\textwidth, trim=0 365 0 355, clip=false]{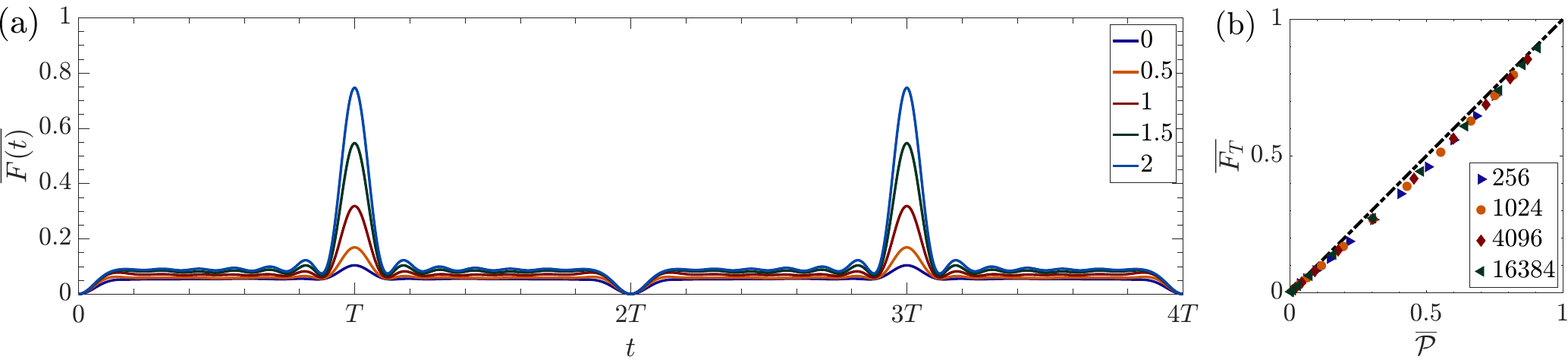}
	\caption[]{{\bf Time evolution of the transfer fidelity} (a) averaged over $10^4$ disordered realizations of $16\times16$ adjacency matrices in Eq.~\eqref{eq_tunable_A} for different $\gamma$. The energy levels of $\adj$ are $\vec{E} = \cbr{1, 2,\dots}$.%, i.e. PST occurs at $T = \pi$ for $\gamma\gg 1$. 
		%============
		(b) the ensemble averaged fidelity at $t = T$ w.r.t. $\overline{\cp}$, the ensemble averaged measure of centrosymmetry (Eq.~\eqref{eq_def_P}) for different system sizes (the dashed curve denotes $\overline{F_T} = \overline{\cp}$).
	}
	\label{fig_5}
\end{figure}
%==========================================================
Hence Eq.~\eqref{eq_state_decomp} implies that the ensemble averaged fidelity at $t = T$ is,
\begin{align}
	\label{eq_F_vs_P}
	\overline{F_T}(\gamma) \approx \abs{\sum_{k = 1}^{N}\frac{\abs{\jk}}{N}}^2 \approx \overline{\cp}(\gamma).
\end{align}
Thus the maximum fidelity of excitation transfer is equivalent to the extent of centrosymmetry as shown in Fig.~\ref{fig_5}(b) for $\adj(\gamma)$ of different sizes.
%=========================================================
\section{Discussions}\label{sec_conclusion}
%===========================================
In this work, our principal goal is to control the transfer fidelity between two vertices of a pre-engineered network in the absence of any environment induced decoherence. We show that the maximum transfer fidelity can be exactly controlled by tuning the extent of centrosymmetry in an adjacency matrix of a network. We construct the Deformed Centrosymmetric Ensemble (DCE) interpolating between the random SC ($H_+$), GOE and SSC ($H_-$) matrices. Various deformed ensembles have been used to study symmetry breaking in small-world networks \cite{Carvalho2}, isospin mixing \cite{Guhr3}, vibrations of crystal block \cite{Carvalho1} and we show that the breaking of centrosymmetry naturally leads to the emergence of DCE. The competition of $H_\pm$ in DCE prompts the system size scaling of the model parameter as $\lambda = N^{-\frac{\gamma}{2}}$. Firstly, we propose a measure of exchange symmetry, $\cp$, and identify the centrosymmetric regime for $\gamma>\gamma_\text{P} \equiv 1$. Secondly, we identify the skew-centrosymmetric regime for $\gamma<\gamma_\text{C} \equiv -1$ based on the measure of chirality, $\cc$. For $-1<\gamma<0$, the fraction of mixed states tends to unity in the thermodynamic limit as $\cp\sim 0$, while $\cc$ becomes extensive. However, $\cc$ does not capture any canonical symmetry implying that $\gamma<\gamma_\text{E}\equiv 0$ is the ergodic regime. Contrarily for $0<\gamma<1$, $\cp$ becomes extensive indicating the breaking of ergodicity with the emergence of correlations due to centrosymmetry.

Next we study the perfect transport in a centrosymmetric network and present two examples with freedom in the choice of (a) the energy levels (b) the energy states of the adjacency matrix. In both the cases, the analytically calculated time evolution of the transfer fidelity is in agreement with the numerical simulations. Finally we show that it is possible to tune the maximum transfer fidelity by breaking centrosymmetry. So from the eigenvectors of DCE we construct an adjacency matrix, $\adj\del{\gamma}$ and we prove that the ensemble averaged maximum fidelity is equivalent to the measure of centrosymmetry, $\cp$.

Our construction can be seen as a realization of a quantum wire allowing tunable transfer of information between the input and the output qubits. The adjacency matrix $\adj\del{\gamma}$ is related to the graph $\graph\del{\gamma}$, where we place $N$ spin-1/2 particles on each vertex, the $i^{th}$ spin has Zeeman energy $\adj_{ii}$ and an isotropic XY interaction, $\adj_{ij}$, with the $j^{th}$ spin. In the underlying Hamiltonian, $\hat{H}_\graph\del{\gamma}$, Z-component of the total angular momentum is conserved and all spins down is an eigenstate. Hence the dynamics of any arbitrary single excitation is confined to the first excitation subspace justifying our study of the evolution of $\ket{1}$ to $\ket{N}$ \cite{Christandl1}. Our construction is applicable to the perfect transfer of many-particle state in a network of indistinguishable non-interacting spinless Fermions obtained by the Jordan-Wigner transformation \cite{Albanese1}. However, it is not clear that the maximum fidelity of multi-particle excitation is same as $\overline{\cp}$ even with the loss of centrosymmetry. Importantly unlike previous studies \cite{Christandl1, Albanese1, Christandl4}, our stochastic approach allows the construction of an infinite number of networks with identical transport properties. Our study proposes a possible mechanism to achieve a controlled transmission in quantum information processing over complex networks.
%===========================================
\begin{acknowledgements}
	A.K.D. is supported by an INSPIRE Fellowship, DST, India.
\end{acknowledgements}
%%********************************************************
%	APPENDIX
%\newpage
\appendix
\renewcommand\thefigure{\thesection.\Roman{figure}}  
\setcounter{figure}{0}
\renewcommand\thetable{\thesection.\Roman{table}}  
\setcounter{table}{0}
\renewcommand{\x}{0.36}
%=========================================================
%	Centrosymmetric Matrix
\section{Symmetric Centrosymmetric (SC) Matrix}\label{sec_rsc}
%=========================================================
%	TABLE containing partition, block diagonal of SC matrix
\begin{table}[t]
	%	\small
	\begin{tabular}{|l|c|c|}
		\hline
		& $N=2m$ & $N=2m+1$ \\ \hline
		partition    & $\begin{pmatrix}
			A& C^T\\
			C& \J A\J 
		\end{pmatrix}$    & $\begin{pmatrix}
			A& \vec{x}& C^T\\
			\vec{x}^T& q& \vec{x}^T\J \\
			C& \J \vec{x}& \J A\J 
		\end{pmatrix}$      \\ \hline
		$Q$          & $\dfrac{1}{\sqrt{2}}\begin{pmatrix}
			-\J Q_1& \J Q_2\\
			Q_1& Q_2
		\end{pmatrix}$    & $\dfrac{1}{\sqrt{2}}\begin{pmatrix}
			-\J Q_1& 0& \J Q_2\\
			0& \sqrt{2}& 0\\
			Q_1& 0& Q_2
		\end{pmatrix}$      \\ \hline
		$H'$ & $\begin{pmatrix}
			\mathcal{A}_1& 0\\
			0& \mathcal{A}_2
		\end{pmatrix}$    & $\begin{pmatrix}
			\mathcal{A}_1& 0& 0\\
			0& q& \sqrt{2}\vec{x}^T\J Q_1\\
			0& \sqrt{2}Q_2^T\J \vec{x}& \mathcal{A}_2
		\end{pmatrix}$      \\ \hline
		$\begin{matrix}
			\textrm{simplest}\\
			H'
		\end{matrix}$ & $\begin{pmatrix}
			A - \J C& 0\\
			0& A + \J C
		\end{pmatrix}$    & $\begin{pmatrix}
			A - \J C& 0& 0\\
			0& q& \sqrt{2}\vec{x}^T\\
			0& \sqrt{2}\vec{x}& A + \J C
		\end{pmatrix}$      \\ \hline
	\end{tabular}
	\caption{\textbf{SC Matrix - Block partitioned structure}: $A, C, \J \in\mathbb{R}^{m\times m}$, where $A = A^T$ and $C = \J C^T\J $, $\vec{x}\in\mathbb{R}^m, q\in\mathbb{R}$. The diagonalizing basis of exchange matrix is denoted by $Q$ expressed in terms of orthonormal matrices $Q_1, Q_2\in\mathbb{R}^{m\times m}$. In the block-diagonal matrix $H' = Q^THQ = \begin{pmatrix}
			B_1& 0\\
			0& B_2
		\end{pmatrix}$, $\mathcal{A}_1 = Q_1^T\J (A - \J C)\J Q_1$ and $\mathcal{A}_2 = Q_2^T\J (A + \J C)\J Q_2$.
	}
	\label{tab_sc_forms}
\end{table}
%=========================================================
An $N\times N$ centrosymmetric matrix, $H$, satisfies the commutation relation $[H, \J ] = 0$ (consequently, $H_{i,j} = H_{N+1-i, N+1-j}\;\forall\; i,j$). In addition to being centrosymmetric, if $H$ is also symmetric (i.e. $H = H^T$), then $H$ is called a Symmetric Centrosymmetric (SC) matrix. SC matrices are also persymmetric \cite{Cantoni1} ($H = \J H^T\J \iff H_{i,j} = H_{N+1-j, N+1-i}$). Any $N\times N$ SC matrix can be partitioned into four non-overlapping blocks, as illustrated in Table~\ref{tab_sc_forms}.
Since $[H,\J ]=0$, we can reduce $H$ in the basis of $\J $ via a similarity transformation $Q^THQ = H'$ such that $H'$ assumes a block diagonal form. General forms of $H'$ are given in Table~\ref{tab_sc_forms}. A particular choice of $Q_1 = -\J , Q_2 = \J $ gives the simplest form of $H'$, containing the diagonal blocks $B_{1,2}$. As $\J $ is conserved for SC matrices, we can separate the eigenspectrum into two symmetry sectors, corresponding to $B_{1,2}$, which are called  odd and even sectors respectively. As shown in Table~\ref{tab_sc_ev}, even and odd sectors give symmetric and anti-symmetric eigenvectors respectively.
%---------------------------------------------------------
%	TABLE containing eigenvectors of SC matrix
\begin{table}[h!]
	%	\centering
	\begin{tabular}{|c|c|c|}
		\hline
		\multicolumn{1}{|c|}{Sector} & \multicolumn{1}{c|}{$\ket{\Psi}_B$}  & \multicolumn{1}{c|}{$\ket{\Psi}_H$} \\ \hline
		\multirow{2}{*}{Odd}  & \multirow{2}{*}{$B_1\vec{u}_j = \lambda_ju_j$} & $\vec{v}_j = \dfrac{1}{\sqrt{2}}\begin{pmatrix}
			\vec{u}_j\\
			-\J \vec{u}_j
		\end{pmatrix}$ \\\cline{3-3} 
		&	& $\vec{v}_j = \dfrac{1}{\sqrt{2}}\begin{pmatrix}
			\vec{u}_j\\
			0\\
			-\J \vec{u}_j
		\end{pmatrix}$                           \\ \hline
		\multirow{2}{*}{Even} & $B_2\vec{y}_j = \alpha_j\vec{y}_j$ & $\vec{w}_j = \dfrac{1}{\sqrt{2}}\begin{pmatrix}
			\vec{y}_j\\
			\J \vec{y}_j
		\end{pmatrix}$ \\ \cline{2-3} 
		& $B_2\begin{pmatrix}
			\gamma_j\\
			\vec{y_j}
		\end{pmatrix} = \alpha_j\begin{pmatrix}
			\gamma_j\\
			\vec{y_j}
		\end{pmatrix}$ & $\vec{w_j} = \dfrac{1}{\sqrt{2}}\begin{pmatrix}
			\vec{y_j}\\
			2\gamma_j\\
			\J \vec{y_j}
		\end{pmatrix}$                          \\ \hline
	\end{tabular}
	\caption{\textbf{SC Matrix:} Eigenvectors from two symmetry sectors, with eigenvalues $\lambda_j$'s, $\alpha_j$'s. $\ket{\Psi}_B$ corresponds to individual blocks, whereas $\ket{\Psi}_H$'s belong to the whole matrix.
	}
	\label{tab_sc_ev}
\end{table}
%=========================================================
%	Skew-Centrosymmetric Matrix
\section{Symmetric Skew-Centrosymmetric (SSC) Matrix}{\label{sec_rss}}
%=========================================================
A skew-centrosymmetric matrix anti-commute with the exchange matrix, i.e. $\cbr{H, \J } = 0$ such that $H_{i,j} = -H_{N+1-i, N+1-j}\;\forall\; i,j$. For skew-centrosymmetric matrices, if $\del{\lambda, \ket{\lambda}}$ is an eigenpair, then so is $\del{-\lambda, \J \ket{\lambda}}$, where $\pm\lambda$ have the same multiplicity \cite{Hill1}. If such a $H$ is symmetric too then we get Symmetric Skew-centrosymmetric (SSC) matrix, which turns out to be skew-persymmetric as well (i.e. $H = -\J H^T\J \iff H_{i,j} = -H_{N+1-j, N+1-i}$) \cite{Cantoni1}. Similar to its centrosymmetric counterpart, SSC matrices also admit  partitioned structures. We can reduce $H$ to off-diagonal blocks in the diagonalizing basis of $\J $ \cite{Abu-Jeib2}. Corresponding structures are summarized in Table~\ref{tab_ss_forms}.
%---------------------------------------------------------
%	TABLE containing partition, block diagonal of SC matrix
\begin{table}[h!]
	%	\centering
	\begin{tabular}{|l|c|c|}
		\hline
		& $N=2m$ & $N=2m+1$ \\ \hline
		partition    & $\begin{pmatrix}
			A& C^T\\
			C& -\J A\J 
		\end{pmatrix}$    & $\begin{pmatrix}
			A& \vec{x}& C^T\\
			\vec{x}^T& 0& -\vec{x}^T\J \\
			C& -\J \vec{x}& -\J A\J 
		\end{pmatrix}$      \\ \hline
		$H'$ & $-\begin{pmatrix}
			0& \mathcal{C}_1\\
			\mathcal{C}_2 &0
		\end{pmatrix}$    & $-\begin{pmatrix}
			0& \sqrt{2}Q_1^T\J \vec{x}& \mathcal{C}_1\\
			\sqrt{2}\vec{x}^T\J Q_1& 0& 0\\
			\mathcal{C}_2& 0& 0
		\end{pmatrix}$      \\ \hline
		$\begin{matrix}
			\textrm{simplest}\\
			H'
		\end{matrix}$ & $\begin{pmatrix}
			0& A - \J C\\
			A + \J C& 0
		\end{pmatrix}$    & $\begin{pmatrix}
			0& \sqrt{2}\vec{x}& A - \J C\\
			\sqrt{2}\vec{x}^T& 0& 0\\
			A + \J C& 0& 0
		\end{pmatrix}$      \\ \hline
	\end{tabular}
	\caption{\textbf{SSC Matrix:} Partition and block diagonal structures. $A, C\in\mathbb{R}^{m\times m}$, where $A = A^T$ and $C = -\J C^T\J $. $\vec{x}\in\mathbb{R}^m$. $Q$ is orthonormal diagonalizing basis of $\J $ (Table~\ref{tab_sc_forms}), where $Q_1, Q_2$ are orthonormal. In $H'$, the blocks are: $\mathcal{C}_1 = Q_1^T\J (A - \J C)\J Q_2$ and $\mathcal{C}_2 = Q_2^T\J (A + \J C)\J Q_1$.}
	\label{tab_ss_forms}
\end{table}
%=========================================================
\section{Shannon entropy and Second moment of DCE}{\label{sec_moment}}
%=========================================================
In the decomposition of a general matrix into a centrosymmetric and skew-centrosymmetric parts, the squared Frobenius norms of the component matrices are
\begin{align*}
	\fnrm{\ch_\pm}^2 &= \Tr{\ch_\pm^2} = \frac{1}{2}\del{\Tr{H^2} \pm \Tr{\del{\J H}^2}}.
\end{align*}
Let, $i' \equiv N+1-i$. Then
\begin{align*}
	(\J H\J)_{ij} &= \sum_{k, l}\J_{ik}H_{kl}\J_{lj}\\
	&= \sum_{k}\delta_{N+1-i, k}H_{kl}\delta_{l, N+1-j} = H_{i'j'}.
\end{align*}
Consequently $\Tr{ (\J H)^2 }$, the second term in $\fnrm{\ch_\pm}^2$ can be written as
\begin{align}
	\label{eq_H_pm_norm}
	\begin{split}
		& \sum_{i}(\J H\J H)_{ii} = \sum_{i, j} (\J H\J)_{ij}H_{ji} = \sum_{i, j} H_{ij}H_{i'j'}\\
		= & \sum_{k = 1}^{N}H_{kk}H_{k'k'} + 2\sum_{i<j}H_{ij}H_{i'j'}\: \del{\because\sum_{i, j} = \sum_{i = j} + \sum_{i \neq j}}\\
		= & 2\sum_{k = 1}^{N/2}H_{kk}H_{k'k'} + 2\del{\sum_{i<j\leq N/2} + \sum_{i\leq N/2<j} + \sum_{N/2<i<j}}H_{ij}H_{i'j'}\\
		= & 2\sum_{k = 1}^{N/2}H_{kk}H_{k'k'} + 4\sum_{i<j\leq N/2}H_{ij}H_{i'j'} + 2\sum_{i\leq N/2 <j}H_{ij}H_{i'j'}\\
		= & 2\underbrace{ \sum_{k = 1}^{N/2}H_{kk}H_{k'k'} }_{N/2 \text{ terms}} + 4 \underbrace{ \sum_{i<j\leq N/2}H_{ij}H_{i'j'} }_{N(N-2)/8 \text{ terms}}\\ &+ 4 \underbrace{ \sum_{i\leq N/2 <j,\; i<j'}H_{ij}H_{i'j'} }_{N(N-2)/8 \text{ terms}} + 2\underbrace{ \sum_{i \leq N/2} H_{ii'}^2 }_{N/2 \text{ terms}}.
	\end{split}
\end{align}
Similarly we can express the squared norm of the matrix and the differential matrix element as
\begin{align}
	\label{eq_dH}
	\begin{split}
		\Tr{H^2} &= \sum_{k = 1}^{N/2}(H_{kk}^2 + H_{k'k'}^2) + 2\sum_{i<j\leq N/2}(H_{ij}^2 + H_{i'j'}^2)\\ &+ 2\sum_{i\leq N/2 <j,\; i<j'}(H_{ij}^2 + H_{i'j'}^2) +  2\sum_{i \leq N/2} H_{ii'}^2\\
		dH &= \prod_{k = 1}^{N/2}dH_{kk}dH_{k'k'} \prod_{\cbr{i, j}}dH_{ij}dH_{i'j'} \prod_{i\leq N/2}dH_{ii'}
	\end{split}
\end{align}
where $\cbr{i, j}\equiv (i<j' \text{ and } i<j\leq \frac{N}{2} \text{ or } i\leq \frac{N}{2} <j)$. Then $\mean{H_{ij}H_{kl}}$, the $2^{nd}$ moment of $H$ can be calculated as
\begin{align}
	\label{eq_2nd_moment}
	\begin{split}
	\mean{H_{ij}H_{kl}} &= \int dH\prob{H}H_{ij}H_{kl}\\
	&= \frac{\delta_{ik'}\delta_{jl'} + \delta_{il'}\delta_{jk'}}{4(\alpha + \beta)}\del{\delta_{ij'} + \frac{(1 - \delta_{ij'})\beta}{\beta - \alpha}}.
	\end{split}
\end{align}
If we take $\alpha = \frac{1}{2}$ and $\lambda = \sqrt{\frac{1-\beta}{1+\beta}}$, then the $2^{nd}$ moment can be expressed as,
\begin{align*}
	\mean{H_{ij}H_{kl}} &= \frac{ \del{1+\lambda^2} \del{\delta_{ik'}\delta_{jl'} + \delta_{il'}\delta_{jk'}} }{4}\\
	&\times \del{\delta_{ij'} + \frac{\lambda^2 - 1}{\lambda^2}\del{1 - \delta_{ij'}}}.
\end{align*}
Similarly the Shannon entropy per degree of freedom is calculated,
\begin{align}
	\label{eq_S_modified}
	\tilde{\shn}(\lambda) = \frac{1}{2}\del{1 + \ln\del{\frac{\pi}{2}\del{\lambda + \frac{1}{\lambda}}}}^{} = \tilde{\shn}\del{\frac{1}{\lambda}}.
\end{align}
For $\lambda = 1$, we have the following expressions
\begin{align}
	\label{eq_GOE}
	\prob{H} = \frac{1}{ 2^{\frac{N}{2}}\pi^{\frac{N(N+1)}{4}} }e^{-\Tr{H^2}},\quad \tilde{S} = \frac{\log\pi + 1}{2}
\end{align}
which are precisely the matrix density and Shannon entropy per degree of freedom of GOE \cite{Hussein2}, as expected for $\lambda = 1$. Due to the symmetry, $\tilde{\shn}(\lambda) = \tilde{\shn}\del{\frac{1}{\lambda}}$, entropy has a minimum at $\lambda = 1$ where $H(\lambda = 1)\in$ GOE. This can be understood from the presence of (anti-)correlation among the elements of (skew-)centrosymmetric matrices and captured by the second moment, $\mean{H_{ij}H_{kl}}$.
%=========================================================
\section{Connection of DCE to DGOE}{\label{sec_dgoe}}
%=========================================================
From Table~\ref{tab_sc_forms}, we know that the orthonormal diagonalizing basis for even ranked exchange matrix is $Q = \dfrac{1}{\sqrt{2}}\begin{pmatrix}
	-\J Q_1& \J Q_2\\
	Q_1& Q_2
\end{pmatrix}$, where $Q_{1, 2}$ are arbitrary orthonormal matrices. Then $\J ' = Q^T \J Q = \mathbb{I}\oplus(-\mathbb{I})$ where $\mathbb{I}$ is the identity matrix. Now we define $\pt = \sum_{i = 1}^{N/2}\ket{i}\bra{i}$ and $\qt = \mathbb{I} - \pt$ where $\ket{i}\equiv$ unit basis vector. As $QQ^T = \mathbb{I}$, $(\J \hgoe \J )' = \J '\hgoe'\J '$, which can be expanded as 
\begin{align*}
	\begin{split}
		&\J'\del{\pt\hgoe'\pt + \pt\hgoe'\qt + \qt\hgoe'\pt + \qt\hgoe'\qt } \J'\\
		= &\pt\hgoe'\pt - \pt\hgoe'\qt - \qt\hgoe'\pt + \qt\hgoe'\qt.
	\end{split}
\end{align*}
Note that $\hgoe'\in$ GOE due to canonical invariance. Then $H_\pm$ from Eq.~\eqref{eq_H_model_J} transforms in the $Q$ basis as
\begin{align*}
	\begin{split}
		H_+' &= \frac{1}{2}\del{\hgoe' + (\J \hgoe \J )'} = \pt \hgoe'\pt + \qt \hgoe' \qt\\
		H_-' &= \frac{1}{2}\del{\hgoe' - (\J \hgoe \J )'} = \pt \hgoe'\qt + \qt \hgoe' \pt.
	\end{split}
\end{align*}
Consequently $H' = Q^THQ = H_+' + \lambda H_-'$ belong to the DGOE \cite{Hussein2}. Thus our matrix model becomes DGOE upon a similarity transformation by $Q$, i.e.~the eigenbasis of the exchange matrix.
%=========================================================
%	TABLES

%=========================================================
%	BIBLIOGRAPHY
%=========================================================

%=========================================================

\begin{thebibliography}{10}
	
	\bibitem{Mandel1}
	O.~Mandel, M.~Greiner, A.~Widera, T.~Rom, T.~W. H{\"a}nsch, and I.~Bloch,
	``Controlled collisions for multi-particle entanglement of optically trapped
	atoms,'' {\em Nature}, vol.~425, no.~6961, pp.~937--940, 2003.
	
	\bibitem{Loss1}
	D.~Loss and D.~P. DiVincenzo, ``Quantum computation with quantum dots,'' {\em
		Phys. Rev. A}, vol.~57, pp.~120--126, Jan 1998.
	
	\bibitem{Gershenfeld1}
	N.~A. Gershenfeld and I.~L. Chuang, ``Bulk spin-resonance quantum
	computation,'' {\em Science}, vol.~275, no.~5298, pp.~350--356, 1997.
	
	\bibitem{Brown1}
	K.~R. Brown, J.~Kim, and C.~Monroe, ``Co-designing a scalable quantum computer
	with trapped atomic ions,'' {\em npj Quantum Information}, vol.~2, no.~1,
	pp.~1--10, 2016.
	
	\bibitem{Karamlou1}
	A.~H. Karamlou, J.~Braum{\"u}ller, Y.~Yanay, A.~Di~Paolo, P.~M. Harrington,
	B.~Kannan, D.~Kim, M.~Kjaergaard, A.~Melville, S.~Muschinske, {\em et~al.},
	``Quantum transport and localization in 1d and 2d tight-binding lattices,''
	{\em npj Quantum Information}, vol.~8, no.~1, pp.~1--8, 2022.
	
	\bibitem{Kane1}
	B.~E. Kane, ``A silicon-based nuclear spin quantum computer,'' {\em Nature},
	vol.~393, no.~6681, pp.~133--137, 1998.
	
	\bibitem{Haselgrove1}
	H.~L. Haselgrove, ``Optimal state encoding for quantum walks and quantum
	communication over spin systems,'' {\em Phys. Rev. A}, vol.~72, p.~062326,
	Dec 2005.
	
	\bibitem{Benjamin1}
	S.~C. Benjamin, ``Quantum computing without local control of qubit-qubit
	interactions,'' {\em Phys. Rev. Lett.}, vol.~88, p.~017904, Dec 2001.
	
	\bibitem{Christandl1}
	M.~Christandl, N.~Datta, A.~Ekert, and A.~J. Landahl, ``Perfect state transfer
	in quantum spin networks,'' {\em Physical review letters}, vol.~92, no.~18,
	p.~187902, 2004.
	
	\bibitem{Karbach1}
	P.~Karbach and J.~Stolze, ``Spin chains as perfect quantum state mirrors,''
	{\em Phys. Rev. A}, vol.~72, p.~030301, Sep 2005.
	
	\bibitem{Shi1}
	T.~Shi, Y.~Li, Z.~Song, and C.-P. Sun, ``Quantum-state transfer via the
	ferromagnetic chain in a spatially modulated field,'' {\em Physical Review
		A}, vol.~71, no.~3, p.~032309, 2005.
	
	\bibitem{Venuti1}
	L.~Campos~Venuti, C.~Degli Esposti~Boschi, and M.~Roncaglia, ``Qubit
	teleportation and transfer across antiferromagnetic spin chains,'' {\em Phys.
		Rev. Lett.}, vol.~99, p.~060401, Aug 2007.
	
	\bibitem{Nikolopoulos1}
	G.~M. Nikolopoulos, D.~Petrosyan, and P.~Lambropoulos, ``Electron wavepacket
	propagation in a chain of coupled quantum dots,'' {\em Journal of Physics:
		Condensed Matter}, vol.~16, pp.~4991--5002, jul 2004.
	
	\bibitem{Albanese1}
	C.~Albanese, M.~Christandl, N.~Datta, and A.~Ekert, ``Mirror inversion of
	quantum states in linear registers,'' {\em Phys. Rev. Lett.}, vol.~93,
	p.~230502, Nov 2004.
	
	\bibitem{Lambert1}
	N.~Lambert, Y.-N. Chen, Y.-C. Cheng, C.-M. Li, G.-Y. Chen, and F.~Nori,
	``Quantum biology,'' {\em Nature Physics}, vol.~9, no.~1, pp.~10--18, 2013.
	
	\bibitem{Scholak1}
	T.~Scholak, F.~de~Melo, T.~Wellens, F.~Mintert, and A.~Buchleitner, ``Efficient
	and coherent excitation transfer across disordered molecular networks,'' {\em
		Phys. Rev. E}, vol.~83, p.~021912, Feb 2011.
	
	\bibitem{Zech1}
	T.~Zech, R.~Mulet, T.~Wellens, and A.~Buchleitner, ``Centrosymmetry enhances
	quantum transport in disordered molecular networks,'' {\em New Journal of
		Physics}, vol.~16, p.~055002, may 2014.
	
	\bibitem{Ortega1}
	A.~Ortega, T.~Stegmann, and L.~Benet, ``Efficient quantum transport in
	disordered interacting many-body networks,'' {\em Phys. Rev. E}, vol.~94,
	p.~042102, Oct 2016.
	
	\bibitem{Gersho1}
	A.~Gersho, ``Adaptive equalization of highly dispersive channels for data
	transmission,'' {\em Bell System Technical Journal}, vol.~48, no.~1,
	pp.~55--70, 1969.
	
	\bibitem{Magee1}
	F.~Magee and J.~Proakis, ``An estimate of the upper bound on error probability
	for maximum-likelihood sequence estimation on channels having a
	finite-duration pulse response (corresp.),'' {\em IEEE Transactions on
		Information Theory}, vol.~19, no.~5, pp.~699--702, 1973.
	
	\bibitem{Datta1}
	L.~Datta and S.~D. Morgera, ``On the reducibility of centrosymmetric
	matrices—applications in engineering problems,'' {\em Circuits, Systems and
		Signal Processing}, vol.~8, no.~1, pp.~71--96, 1989.
	
	\bibitem{Gaudreau1}
	{Gaudreau, Philippe} and {Safouhi, Hassan}, ``Centrosymmetric matrices in the
	sinc collocation method for sturm-liouville problems,'' {\em EPJ Web of
		Conferences}, vol.~108, p.~01004, 2016.
	
	\bibitem{Collar1}
	A.~R. Collar, ``{On Centrosymmetric and Centroskew Matrices},'' {\em The
		Quarterly Journal of Mechanics and Applied Mathematics}, vol.~15,
	pp.~265--281, 08 1962.
	
	\bibitem{Cantoni1}
	A.~Cantoni and P.~Butler, ``Eigenvalues and eigenvectors of symmetric
	centrosymmetric matrices,'' {\em Linear Algebra and its Applications},
	vol.~13, no.~3, pp.~275 -- 288, 1976.
	
	\bibitem{Abu-Jeib1}
	I.~T. Abu-Jeib, ``Centrosymmetric and skew-centrosymmetric matrices and regular
	magic squares,'' {\em New Zealand J. Math}, vol.~33, no.~2, pp.~105--112,
	2004.
	
	\bibitem{Beenakker1}
	C.~W.~J. Beenakker, ``Random-matrix theory of majorana fermions and topological
	superconductors,'' {\em Rev. Mod. Phys.}, vol.~87, pp.~1037--1066, Sep 2015.
	
	\bibitem{Schneider1}
	B.~Schneider, J.~C. Halimeh, and M.~Punk, ``Projective symmetry group
	classification of chiral $z_2$ spin liquids on the pyrochlore lattice:
	application to the spin-1/2 xxz heisenberg model,'' {\em Phys. Rev. B}, vol.~105,
	p.~125122, Mar 2022.
	
	\bibitem{Verbaarschot1}
	J.~Verbaarschot and T.~Wettig, ``Random matrix theoery and chiral symmetry in
	qcd,'' {\em Annual Review of Nuclear and Particle Science}, vol.~50, no.~1,
	pp.~343--410, 2000.
	
	\bibitem{Rehemanjiang1}
	A.~Rehemanjiang, M.~Richter, U.~Kuhl, and H.-J. St\"ockmann, ``Microwave
	realization of the chiral orthogonal, unitary, and symplectic ensembles,''
	{\em Phys. Rev. Lett.}, vol.~124, p.~116801, Mar 2020.
	
	\bibitem{Hussein2}
	M.~S. Hussein and M.~P. Pato, ``Description of chaos-order transition with
	random matrices within the maximum entropy principle,'' {\em Phys. Rev.
		Lett.}, vol.~70, pp.~1089--1092, Feb 1993.
	
	\bibitem{Das3}
	A.~K. Das and A.~Ghosh, ``Chaos due to symmetry-breaking in deformed poisson
	ensemble,'' {\em Journal of Statistical Mechanics: Theory and Experiment},
	vol.~2022, no.~6, p.~063101, 2022.
	
	\bibitem{Mehta1}
	M.~Mehta, {\em Random Matrices}.
	\newblock Pure and Applied Mathematics, Elsevier Science, 2004.
	
	\bibitem{Kravtsov1}
	V.~E. Kravtsov, I.~M. Khaymovich, E.~Cuevas, and M.~Amini, ``A random matrix
	model with localization and ergodic transitions,'' {\em New Journal of
		Physics}, vol.~17, p.~122002, dec 2015.
	
	\bibitem{Das1}
	A.~K. Das and A.~Ghosh, ``Eigenvalue statistics for generalized symmetric and
	hermitian matrices,'' {\em Journal of Physics A: Mathematical and
		Theoretical}, vol.~52, p.~395001, sep 2019.
	
	\bibitem{Berry5}
	M.~V. Berry, ``Regular and irregular semiclassical wavefunctions,'' {\em
		Journal of Physics A: Mathematical and General}, vol.~10, pp.~2083--2091, dec
	1977.
	
	\bibitem{Hill1}
	R.~D. Hill, R.~G. Bates, and S.~R. Waters, ``On centrohermitian matrices,''
	{\em SIAM Journal on Matrix Analysis and Applications}, vol.~11, no.~1,
	pp.~128--133, 1990.
	
	\bibitem{Das2}
	A.~K. Das and A.~Ghosh, ``Nonergodic extended states in the
	$\ensuremath{\beta}$ ensemble,'' {\em Phys. Rev. E}, vol.~105, p.~054121, May
	2022.
	
	\bibitem{Christandl4}
	M.~Christandl, L.~Vinet, and A.~Zhedanov, ``Analytic next-to-nearest-neighbor x
	x models with perfect state transfer and fractional revival,'' {\em Physical
		Review A}, vol.~96, no.~3, p.~032335, 2017.
	
	\bibitem{Song1}
	Z.~Song and C.~Sun, ``Quantum information storage and state transfer based on
	spin systems,'' {\em Low Temperature Physics}, vol.~31, no.~8, pp.~686--694,
	2005.
	
	\bibitem{Yung1}
	M.-H. Yung and S.~Bose, ``Perfect state transfer, effective gates, and
	entanglement generation in engineered bosonic and fermionic networks,'' {\em
		Physical Review A}, vol.~71, no.~3, p.~032310, 2005.
	
	\bibitem{Kostak1}
	V.~Kostak, G.~M. Nikolopoulos, and I.~Jex, ``Perfect state transfer in networks
	of arbitrary topology and coupling configuration,'' {\em Phys. Rev. A},
	vol.~75, p.~042319, Apr 2007.
	
	\bibitem{Luca1}
	A.~De~Luca, B.~L. Altshuler, V.~E. Kravtsov, and A.~Scardicchio, ``Anderson
	localization on the bethe lattice: Nonergodicity of extended states,'' {\em
		Phys. Rev. Lett.}, vol.~113, p.~046806, Jul 2014.
	
	\bibitem{Carvalho2}
	J.~X. de~Carvalho, S.~Jalan, and M.~S. Hussein, ``Deformed
	gaussian-orthogonal-ensemble description of small-world networks,'' {\em
		Phys. Rev. E}, vol.~79, p.~056222, May 2009.
	
	\bibitem{Guhr3}
	T.~Guhr and H.~Weidenm{\"u}ller, ``Isospin mixing and spectral fluctuation
	properties,'' {\em Annals of Physics}, vol.~199, no.~2, pp.~412--446, 1990.
	
	\bibitem{Carvalho1}
	J.~X. de~Carvalho, M.~S. Hussein, M.~P. Pato, and A.~J. Sargeant,
	``Symmetry-breaking study with deformed ensembles,'' {\em Phys. Rev. E},
	vol.~76, p.~066212, Dec 2007.
	
	\bibitem{Cantoni1}
	A.~Cantoni and P.~Butler, ``Eigenvalues and eigenvectors of symmetric
	centrosymmetric matrices,'' {\em Linear Algebra and its Applications},
	vol.~13, no.~3, pp.~275 -- 288, 1976.
	
	\bibitem{Hill1}
	R.~D. Hill, R.~G. Bates, and S.~R. Waters, ``On centrohermitian matrices,''
	{\em SIAM Journal on Matrix Analysis and Applications}, vol.~11, no.~1,
	pp.~128--133, 1990.
	
	\bibitem{Abu-Jeib2}
	I.~T. Abu-Jeib, ``Centrosymmetric matrices: properties and an alternative
	approach,'' {\em Canadian Applied Mathematics Quarterly}, vol.~10, no.~4,
	pp.~429--445, 2002.
	
\end{thebibliography}
\end{document}